\documentclass[10pt,twocolumn,sort&compress,5p]{elsarticle}

\usepackage{lineno,hyperref}

\usepackage{graphics}
\usepackage{amssymb}
\usepackage{amsmath}
\usepackage{color} % ACTIVATES COLORS

\modulolinenumbers[5]

\journal{Physics Letters B}

\newcommand{\tev}{\text{ TeV}} % Notice the space
\newcommand{\gev}{\text{ GeV}} % Notice the space
\begin{document}

\begin{frontmatter}

%%%%%%%%%%     TITLE     %%%%%%%%%%
\title{\mathversion{bold}Standard Model Extension with Flipped Generations}
%%%%%%%%%% AUTHORS     %%%%%%%%%%
\author[UNotreDame,TsinghuaU]{Carlos Alvarado}
\ead{arcarlos00@gmail.com}
\author[UColima]{Alfredo Aranda}
\ead{fefo@ucol.mx}

%%%%%%%%%%     ADDRESSES     %%%%%%%%%%
\address[UNotreDame]{Department of Physics, University of Notre Dame, 225 Nieuwland Hall, Notre Dame, Indiana 46556, USA.}
\address[TsinghuaU]{Department of Physics, Tsinghua University, Beijing 100084, China}
\address[UColima]{Facultad de Ciencias, CUICBAS, Universidad de Colima, 28040 Colima, M\'exico}

%%%%%%%%%%          ABSTRACT     %%%%%%%%%%
\begin{abstract}
An extension of the Standard Model is presented that leads to the possible existence of new gauge bosons with masses in the range of a few TeV. Due to the fact that their couplings to Standard Model fermions are strongly suppressed, it is possible for them to be {\it hidden} from current searches. The model contains additional generations of fermions with quantum numbers resembling those of the Standard Model fermion generations but with a twist: their charge assignments are such that their electric charges and chiralities are {\it flipped} with respect to those of their corresponding Standard Model counterparts. This feature provides a way to obtain potential dark matter candidates and the interesting possibility for a Lepton number conserving dimension-five operator for Dirac neutrino masses. The model implications associated to electroweak precision parameters, flavor changing neutral currents, and diphoton rate contributions are briefly discussed. The general assumptions of this set up are also used to sketch a couple of variants of the model with peculiar features that could motivate further study.
\end{abstract}

%%%%%%%%%%     KEYWORDS     %%%%%%%%%%

\end{frontmatter}

%\linenumbers

\section{Introduction}
\label{sec:intro}

The search for new phenomena at high energy scales has inspired many different - sometimes complementary - proposals for new physics. From adding extra scalars and/or fermion generations to the Standard Model (SM), extending its gauge sector, adding flavor symmetries, all the way to considering complete grand unified models where most of the salient features of contemporary particle physics are addressed. So far no evidence of any such extensions has showed up in any experiment. At the same time, solutions to several interesting and important questions remain elusive. Some of them are: Where does neutrino mass come from? Can the wild fermion mass spectrum be explained? Why matter over antimatter? What is dark matter?

In this paper we present a set up where additional fermion \textquotedblleft generations" are introduced to the SM such that they are consistent with current experimental bounds and might contribute to interesting phenomenology at the Large Hadron Collider (LHC) and/or provide dark matter candidates. Several old and new ideas related to additional generations of fermions exist in the literature, ranging from studies related to the currently disfavored fourth fermion generation \cite{Djouadi:2012ae,Vysotsky:2013gfa,Lenz:2013iha} to frameworks that involve the existence of several generations of so-called {\it mirror fermions} \cite{Holdom:2009rf,Abdullah:2016avr,Chakdar:2015sra,Arhrib:2006pm,Chanowitz:2012nk,Bar-Shalom:2016ehq,Bulava:2013ep,Geller:2012tg,Frampton:1999xi}. Implementations of this kind are sometimes embedded in Left-Right models \cite{PhysRevD.11.2558,PhysRevD.17.2462,PhysRevD.44.837,PhysRevD.57.4174,PhysRevLett.110.151802,PhysRevD.36.274,PhysRevD.36.878,Aranda:2010sr,Kownacki:2017sqn,Barenboim:2001vu,Brahmachari:2003wv,BANDYOPADHYAY2017206,Chakrabortty:2016wkl,REIG201735,Lindner:1996tf}. An important ingredient or guiding principle in most of the previous extensions, from the theoretical point of view, has always been the possibility of unification. Namely, extra fermion multiplets are considered that can in principle be conceived from a grand unified point of view. This is important because it works as a guide and/or constraint that provides theoretical appeal. Our set up is motivated by those ideas but its main difference basically resides in giving up the constraint (or condition) that the additional fields might be related to some sort of unification. That is certainly a major departure, for it might seem unmotivated by some readers, but it is a venue worth exploring that might contain some interesting consequences.

To do so, we first consider an extended gauge sector consisting of an additional $SU(2)_{hid}$ gauge symmetry {\it hidden} to the SM fermions and under which new fermions (including right-handed neutrinos) will transform non-trivially. The general (gauge) structure of the model is given by
\begin{eqnarray}
\label{gaugegroup}
SU(3)_C \times SU(2)_w \times SU(2)_{hid} \times U(1)_X \ .
\end{eqnarray}

The set up assumes a sequential spontaneous symmetry breaking (SSB) triggered by the the vacuum expectation values (vevs) of scalars in the following way
\begin{eqnarray}
\label{SSB}
\nonumber
SU(3)_C \times  SU(2)_w \times SU(2)_{hid} \times & U(1)_X & \\ \nonumber \longrightarrow \ \ 
SU(3)_C \times SU(2)_w \times & U(1)_Y & \\ 
\longrightarrow \ \ 
SU(3)_C \times & U(1)_{em}.&
\end{eqnarray}
The two scalars that mainly drive this breaking pattern have transformation properties given by $H_{hid} \sim ({\bf{1,1,2}},-1/2)$ and $H \sim ({\bf{1,2,1}},+1/2)$ under the underlying gauge group, and  where $SU(3)_C \times U(1)_{em}$ correspond to the usual SM unbroken gauge groups. The electric charge generator is given by $Q = T^3_w +T^3_{hid} + X = T^3_w + Y$, with $Y \equiv T^3_{hid} + X$, where $T^3_w$ and $T^3_{hidd}$ are the diagonal $SU(2)_w$ and $SU(2)_{hid}$ generators. Note that with the SM fields chosen as singlets under $SU(2)_{hid}$ their $X$-charge is their hypercharge. By construction, we assume hierarchical vevs: $\langle H_{hid}\rangle \gg \langle H\rangle$. 

In addition to the gauge extension, our set up uses the following convention (choice): for any field that we add that has a counterpart within the SM (for example $H_{hid}$ has counterpart $H$), its $X$-charge (and chirality in the case of fermions) will be {\it flipped} (that is why the $X$-charge of $H_{hid}$ is  the negative of the charge of $H$ in the paragraph above). If additional fields that do not have a counterpart in the SM are needed, they will be $U(1)_X$-neutral and their electrical charges will be determined exclusively by their $SU(2)_w \times SU(2)_{hid}$ transformation properties. Although our extended SM gauge group includes another SU(2) factor, and despite the resemblance to Left-Right scenarios in the literature, we would like to stress that ours \textit{is not} one of them: right-handed fermions \textit{are not} promoted to doublets of $SU(2)_{hid}$. Furthermore, $H$ and $H_{hid}$ \textit{do not} transform into each other under the parity operation $\mathcal{P}$. In other words, we won't attempt to accomodate $\mathcal{P}$ (or its breaking) at high energies.

Back to our set up, the additional fields in the model consist of three  {\it flipped} fermion generations and one $X$-neutral scalar bidoublet. In the rest of the paper we present a section with a detailed description of the model, another one with its salient phenomenological features, and we end with our conclusions. We have also included two short appendices with details pertaining to anomaly cancellation and the scalar potential of the model.

\section{The model}
\label{sec:model}
As discussed in the introduction, the gauge symmetry of the model is $SU(3)_C \times SU(2)_w \times SU(2)_{hid} \times U(1)_X$. It consists of the usual three generations of SM fermions \textemdash taken to be singlets of $SU(2)_{hid}$ \textemdash and three additional {\it flipped} generations (see Table \ref{table:field-content}).

\begin{table}
	\begin{tabular}{|c|c|c|c|c|}
	\hline
	{\rm{Field}} & $SU(3)_C$ & $SU(2)_w$ & $SU(2)_{hid}$ & $U(1)_X$  	\\
	\hline \hline
	$Q_L$ & \bf 3 & \bf 2 & \bf 1 & $+1/6$ \\
	$Q'_R$& \bf 3 & \bf 1 & \bf 2 & $-1/6$ \\
	$U_R$& \bf 3 & \bf 1 & \bf 1 & $+2/3$ \\
	$U'_L$& \bf 3 & \bf 1 & \bf 1 & $-2/3$ \\
	$D_R$& \bf 3 & \bf 1 & \bf 1 & $-1/3$ \\
	$D'_L$& \bf 3 & \bf 1 & \bf 1 & $+1/3$ \\
	\hline \hline
	$L$ & \bf 1 & \bf 2 & \bf 1 & $-1/2$ \\
	$R'$& \bf 1 & \bf 1 & \bf 2 & $+1/2$ \\
	$E_R$& \bf 1 & \bf 1 & \bf 1 & $-1$ \\
	$E'_L$& \bf 1 & \bf 1 & \bf 1 & $1$ \\
	\hline \hline
	$H$ & \bf 1 & \bf 2 & \bf 1 & $+1/2$ \\
	$H_{hid}$ & \bf 1 & \bf 1 & \bf 2 & $-1/2$ \\
	$\mathcal{B}$ & \bf 1 & \bf 2 & \bf 2 & $0$ \\
	\hline
	\end{tabular}
	\caption{Transformation properties of the fermion and scalar field content of the model. Family indices are not shown and three generations for all fermion fields are included. Fermion fields not present in the SM are primed.}
	\label{table:field-content}
\end{table}

The gauge bosons present in the model after the SSB in \eqref{SSB} consist of the usual eight massless gluons, the photon $\gamma$, the $Z$ and $W^{\pm}$ of the SM, plus additional $Z'$ and $W'^{\pm}$ gauge bosons. Due to the breaking pattern, the photon and $Z$ boson are linear combinations of the gauge boson associated to $U(1)_X$ and those associated to the diagonal generators of both $SU(2)_w$ and $SU(2)_{hid}$, while the $W^{\pm}$ bosons are linear combinations of the gauge bosons associated to their off-diagonal counterparts.

As discussed earlier, the charge assignments of the new fields are {\it flipped} with respect to those of the SM fields. This set up has some interesting consequences: note for instance that the $SU(2)_{hid}$ lepton doublet $R'$ has its electromagnetic neutral component in the {\it lower} position, i.e. flipped with respect to the $SU(2)_w$ lepton doublet $L$. For the quark sector the same situation is true and thus the labels $U$ and $D$ lose their \textquotedblleft purpose" when used in the flipped sector. To make this more evident we express them in the usual $SU(2)$ notation:
\begin{eqnarray}
\label{fermion-doublets}
L & \equiv & \left(\begin{array}{c}
\nu_L \\ E_L
\end{array}\right), \ \
R'\equiv
\left(\begin{array}{c}
E'_R \\ \nu_R
\end{array}\right), \\
Q_L & \equiv & \left(\begin{array}{c}
U_L \\ D_L
\end{array}\right), \ \
Q'_R\equiv
\left(\begin{array}{c}
D'_R \\ U'_R
\end{array}\right).
\end{eqnarray}

A brief observation regarding the particle content in Table~\ref{table:field-content} is that since the $U(1)_X$ charges of the additional fields correspond to those of the SM (flipped), and since complete generations are introduced, the usual gauge anomaly conditions are automatically satisfied for an arbitrary number of flipped generations. This is explicitly shown in \ref{app:anomI}.

The transformation and charge assignments introduced in Table~\ref{table:field-content} forbid any mixing between SM and flipped fermions. Furthermore, barring the presence of the bidoublet $\mathcal{B}$, the only {\it communication} between the SM and the hidden sector is through the photon/gluon exchange and the quartic portal-like interaction in the scalar potential $H^{\dagger}HH_{hid}^{\dagger}H_{hid}$. This has as immediate consequence the presence of electrically charged stable particles, since none of the interactions that communicate both sectors can transfer electric charge. This is the reason for the incorporation of $\mathcal{B}$, as is described later in this section.
The Yukawa sector of the model is given by\footnote{The minus sign compensates that of the neutral component of $\widetilde{H}_{hid}$.}
\begin{eqnarray}
	\label{eq:yukawa}
	\nonumber
	{\cal L}_{Y} &=& {\cal Y}^{U}\overline{Q_L}\widetilde{H}U_R+{\cal Y}^{D}\overline{Q_L} H D_{R}+{\cal Y}^{E}\overline{L}H E_{R} \\ 
	&-& \nonumber
	{\cal Z}^{U}\overline{Q'_{R}}\widetilde{H}_{hid}U'_{L}+{\cal Z}^{D}\overline{Q'_{R}}H_{hid}D'_{L} \\ 
	&+& {\cal Z}^{E}\overline{R'}H_{hid}E'_{L} + \text{H.c.}~,
\end{eqnarray}
where $\widetilde{H}\equiv i\sigma_2 H^*$, $\widetilde{H}_{hid}\equiv i\sigma_2 H_{hid}^*$ and the ${\cal Y}^{U,D,E}$ and ${\cal Z}^{U,D,E}$ are Yukawa matrices. Observe that the bidoublet ${\mathcal B}$ does not participate in the Yukawa sector and thus does not mix flipped and SM fermions.

Neutrinos are massless at tree level in this set up, including the right-handed ones present in the flipped sector. Note, however, that Dirac neutrino masses can be obtained through the Lepton number conserving, dimension-five operator (see~\cite{Ma:2016mwh} for possible realizations of this type of operator)
\begin{equation}
	\label{eq:dim5nu}
m^D_{\nu} \sim \frac{1}{\Lambda}(\overline{L}\widetilde{H})(\widetilde{H}_{hid}^{\dagger}R')~.
\end{equation}
 It is interesting to note that if neutrinos are Dirac and Lepton number is conserved, the neutrino masses in this model are related to the vevs of the doublet scalars and to the (undetermined) energy cutoff $\Lambda$, which is associated to the scale at which this set up would cease to be effective and is presumably much higher than the scale $v_{hid}$ of the first stage of the sequential SSB.

Let's now turn to the presence of electrically charged states among the extra field content. Such states, if stable, would have been detected already~\footnote{See, however, the mass-to-charge limits from the interaction with galactic magnetic fields \cite{Chuzhoy:2008zy}.}, which leaves us with no choice other than to ensure that they transfer their electric charge back to SM charged states: this is the role of ${\mathcal B}$.  We need interaction vertices that {\it connect} the $W'^{\pm}, q'$, and $\ell'$ to electrically charged SM states. Before showing how ${\mathcal B}$ provides such an interaction, we describe the new electrically charged states of the model. First note that the new flipped quarks, if not allowed to decay to SM quarks, will form new hadron states similar to those in the SM. We assume that in the hidden sector a new flipped and massive neutron-proton pair $n'- p'$ is generated ($p'$ with negative charge) such that their masses satisfy $m_{n'} < m_{p'}$ and the beta-like process $p' \to n'  e'^{-}\nu_R$ is allowed. In this way the (neutral) $n'$ is stable and the only electrically charged new particle is the lightest\footnote{The new charged leptons $\ell'$ could be mass-degenerate, but for simplicity consider the case where $e'$ is the lightest one.} flipped lepton $e'^{-}$ .

In order to make $e'^{\pm}$ unstable we could try to couple it {\it directly} to the SM charged leptons. Suppose a $SU(2)_w\times SU(2)_{hid}$ singlet scalar field $\varphi \sim(\boldsymbol{1},\boldsymbol{1},\boldsymbol{1})_{-2}$ is introduced (against our convention/choice of adding only $X$-neutral fields when no SM counterpart exists). This enables a Yukawa $\overline{E_{R}}E'_{L}\varphi$ and induces the decay $e'^{-}\rightarrow e^{+}\varphi^{--}$. However, as opposed to scenarios where  doubly-charged scalars originate from large $SU(2)_{w}$ multiplets, decay channels of $\varphi^{--}$ to $W^{-}$ plus singly-charged scalars or to a $W^{-}W^{-}$ pair are not available due to the singlet nature of $\varphi$ under that group. Therefore, the doubly-charged singlet scalar would be stable and our problem would persist.

\begin{figure}[h!]
\centering
\includegraphics[scale=0.6]{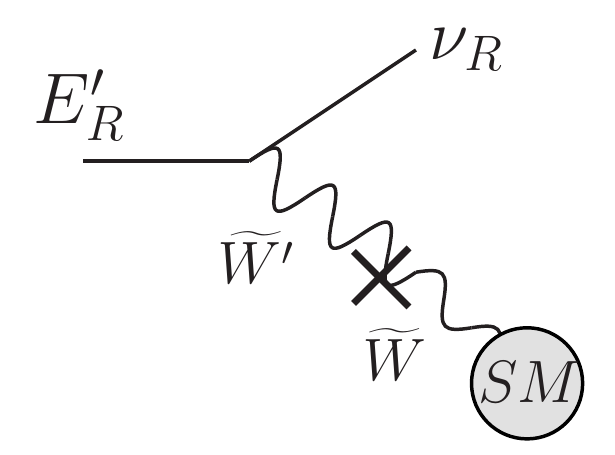}
\caption{Generic decay of an electrically charged, color-singlet hidden fermion into SM states. This is possible through the mixing of the interaction-basis $\widetilde{W}$ and $\widetilde{W}'$ once both bidoublet neutral components develop a vev.}
\label{fig:ellPrimeToSM}
\end{figure}

Our solution is the introduction of the following $U(1)_X$ neutral bidoublet:
\begin{equation}
\mathcal{B}=
\left(\begin{array}{cc}
b_{1}^{0} & b_{2}^{+} \\
b_{1}^{-} & b_{2}^{0}
\end{array}\right)~.
\label{eq:bidoublet}
\end{equation}
As mentioned above, this field does not participate in the Yukawa sector and neither in the generation of neutrino mass. Although flipped-SM fermion mixing through $\mathcal{B}$ is forbidden, this bidoublet introduces charged gauge boson mixing ($\propto \langle b_{1}^{0}\rangle \langle b_{2}^{0}\rangle$) when both of its neutral components develop a non-vanishing vev, opening up an interaction that deploys the electric charge of $e'$ into the SM. This is illustrated in Fig.~\ref{fig:ellPrimeToSM}, where a generic process is shown in the interaction basis. Interestingly enough, $\mathcal{B}$ does not introduce unsuppressed neutrino mass operators either: at renormalizable level neither of $\overline{L}\mathcal{B}R'$ of $\overline{L}\widetilde{\mathcal{B}}R'$ is permitted by gauge invariance.
Beyond renormalizable level the dim-5 operator $\overline{L}\mathcal{B}\mathcal{B}^{\dag}R'$ and other combinations of $\mathcal{B}$ and $\widetilde{\mathcal{B}}$ with two fermion doublets fail to be gauge invariant because $X(\overline{L})+X(R')=+1$. Finally, all dimension-5 combinations of the fermion bilinears $\overline{L}R'$, $\overline{Q'_{R}}Q_{L}$ with  two scalar doublets, 
\begin{align*}
\overline{Q_{L}}HH_{hid}^{\dag}Q_{R},&~~~\overline{Q_{L}}H\widetilde{H}_{hid}^{\dag}Q_{R} \\
\overline{L}HH_{hid}^{\dag}R',&~~~\overline{L}H\widetilde{H}_{hid}^{\dag}R',~~~\overline{L}\widetilde{H}H_{hid}^{\dag}R',
\end{align*}
are ruled out, \textit{except} for the operator $\overline{L}\widetilde{H}\widetilde{H}_{hid}^{\dag}R'$, which generates Dirac neutrino mass terms.

In summary, the {\it role} of the bidoublet $\mathcal{B}$ in the model is to prevent the presence of new stable charged states while being compatible with suppressed, higher-dimensional Dirac neutrino mass operators.

As a side remark please note that there can be additional solutions besides introducing $\mathcal{B}$. For example, going one step further and again relaxing our convention (choice) of setting extra fields to be neutral with respect to $U(1)_X$, we could introduce a bidoublet with $X=+1$. Such a field would allow charged flipped fermions to decay to SM fields: as in the case of the $X=-2$ singlet discussed above, it contains a doubly-charged scalar but also a singly-charged one, thus evading the problem of that case. The downside is that a contribution to neutrino masses would exist at renormalizable level, taking over the operator $\overline{L}\widetilde{H}\widetilde{H}_{hid}^{\dag}R'$ and one would need to find ways to suppress it. In this work we stick to the case of the neutral bidoublet and will defer the $X=+1$ bidoublet possibility to a future publication.

To close this section we present some details of the scalar sector. The renormalizable, gauge invariant potential for the three scalar fields $H$, $H_{hid}$ and ${\mathcal B}$ is given by
\begin{align}
V
&= -\mu^{2}H^{\dag}H+\lambda(H^{\dag}H)^{2}-\mu'^{2}H_{hid}^{\dag}H_{hid}  \notag \\
&+ \lambda'(H_{hid}^{\dag}H_{hid})^{2}+\lambda_{HH}H^{\dag}HH_{hid}^{\dag}H_{hid} \notag \\
&+V_{\mathcal{B}}(H,H_{hid}), \label{eq:Vshort}
\end{align}
where $V_{\mathcal{B}}(H,H_{hid})$ stands for the free potential of the bidoublet plus its interactions with the doublets (shown fully in \ref{app:VHiggs}). Both neutral components of the doublets acquire a vev, $H\rightarrow (0,v_{w}/\sqrt{2})^{T}$, $H_{hid}\rightarrow (v_{hid}/\sqrt{2},0)^{T}$, and so do the neutral components of the bidoublet, $b_{1,2}\rightarrow v_{b1,b2}/\sqrt{2}$. After the two stages of symmetry breaking, and upon minimization (refer to \ref{app:VHiggs}) the sixteen starting degrees of freedom arrange themselves into the following interaction bases: four $CP$-even scalars, $\{ h^{0},h_{hid}^{0},b_{1r}^{0},b_{2r}^{0}\}$, four pseudoscalars $\{ a^{0},a_{hid}^{0},b_{1i}^{0},b_{2i}^{0}\}$ and four charged scalars $\{ H^{+},H_{hid}^{+},b_{1}^{+},b_{2}^{+} \}$. After diagonalization, the neutral physical states left are a light SM-like Higgs, two mostly-bidoublet true scalars $H_{2,3}^{0}$, a heavy scalar made mainly of $H_{hid}^{0}$, two pseudoscalars $A_{1,2}^{0}$ from $\mathcal{B}$, and two charged states $H_{1,2}^{\pm}$ predominantly in the $b_{1,2}^{\pm}$ direction. Such compositions are a result of the vev hierarchy, $v_{b1,2}\ll v_{w} \ll v_{hid}$, with the actual values set by the chosen benchmark parameters. The remaining six degrees of freedom become the longitudinal modes of $W$, $W'$, $Z$ and $Z'$.

Strictly, the bidoublet vevs participate in \textit{both} symmetry breaking stages, however their relative size compared to $v_{w}$ and $v_{hid}$ renders their contribution merely into a correcting effect. Then, the total electroweak vev is defined by $v\equiv \sqrt{v_{w}^{2}+v_{b1}^{2}+v_{b2}^{2}}=246\gev$. It turns useful to change variables according to $v_{b1}\equiv v_{b}\cos{\beta_{b}}$, $v_{b2}\equiv v_{b}\sin{\beta_{b}}$ and $v_{b}\equiv v\cos{\beta}$, $v_{w}\equiv v\sin{\beta}$. Notice that a large $\tan{\beta}$ regime leaves $v_{w}\approx v$, in accordance with $\langle H \rangle$ being the dominant contributor to electroweak symmetry breaking.

\section{Phenomenology highlights}
\label{sec:pheno}
Our main intention in this paper is to present a model with the characteristics mentioned above, namely (i) no additional, stable charged fermions, and (ii) neutrino masses occuring only as dim-5 operators.  A complete phenomenological analysis of the model is beyond the scope of this paper and we intend to present it in a later publication, however there are a few notable phenomenological aspects that we want to comment on this section with the purpose of further motivating our set up.

First, the model is free of tree-level flavor changing neutral currents (FCNCs) by construction. It was already specified in Sec.~\ref{sec:model} that the SM fermions form the usual Yukawa terms with $H$ and that these are the only ones permitted at tree-level. This holds true even after the addition of the bidoublet $\mathcal{B}$, as discussed earlier. That FCNCs are absent at tree-level can be seen directly as follows: starting with Eq.~(\ref{eq:yukawa}), after rotating the $D$-quarks and $CP$-even scalars into physical fields $\widehat{D}$ and $\widehat{h}_{k}^{0}$ we are left with the scalar-fermion-fermion interactions
\begin{equation}
\dfrac{\widehat{\mathcal{Y}}_{i}^{D}}{\sqrt{2}}\overline{\widehat{D}_{Li}}\widehat{D}_{Ri}\biggl( (Z_{H})_{1,1}\widehat{h}_{1}^{0}+\sum_{k>1}^{4}(Z_{H})_{k,1}\widehat{h}_{k}^{0} \biggr)+\text{H.c.}
\end{equation}
Here $\widehat{\mathcal{Y}}_{i}^{D}=\widehat{\mathcal{Y}}_{ij}^{D}\delta_{ij}$ are the diagonal Yukawas, $Z_H$ denotes the orthogonal rotation matrix that diagonalized the scalar mixing matrix, and the $\widehat{h}_{k}^{0}$ are labelled in order of increasing mass. As long as the $\widehat{h}_{k>1}^{0}$ are made heavier than $h_{\text{SM}}\equiv \widehat{h}_{1}^{0}$ (the lightest, SM-like Higgs) it holds that $(Z_{H})_{1,1}\approx 1 \gg (Z_{H})_{1,k}$ and they \textit{do} couple to SM fermion pairs but with a $Z_{H}$ suppression factor. Since the same statement is true for up-type quarks and leptons, these sets of Yukawa couplings will be flavor-diagonal and will not generate tree-level FCNCs in accordance with the Glashow-Weinberg condition \cite{PhysRevD.15.1958}. An analysis at loop-level is currently under preparation for this model (and for other variants mentioned at the end of this work) but due to the small mixing in the scalar and gauge sectors, contributions to flavor changing loop-level processes are generally expected to be also under control and not to impose severe bounds on the parameters of the model.

\begin{figure}
\centering
\includegraphics[scale=0.34]{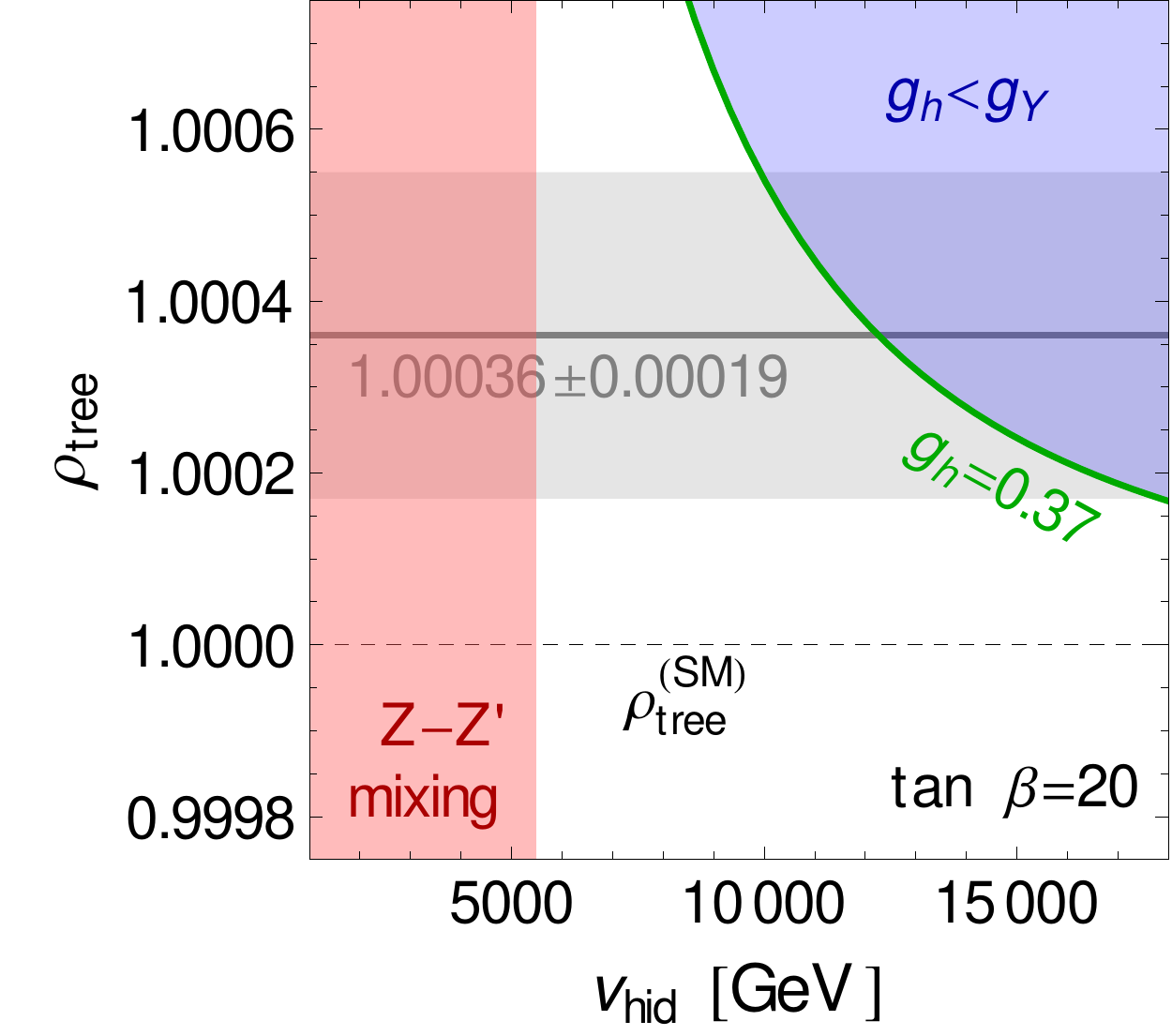}
\includegraphics[scale=0.34]{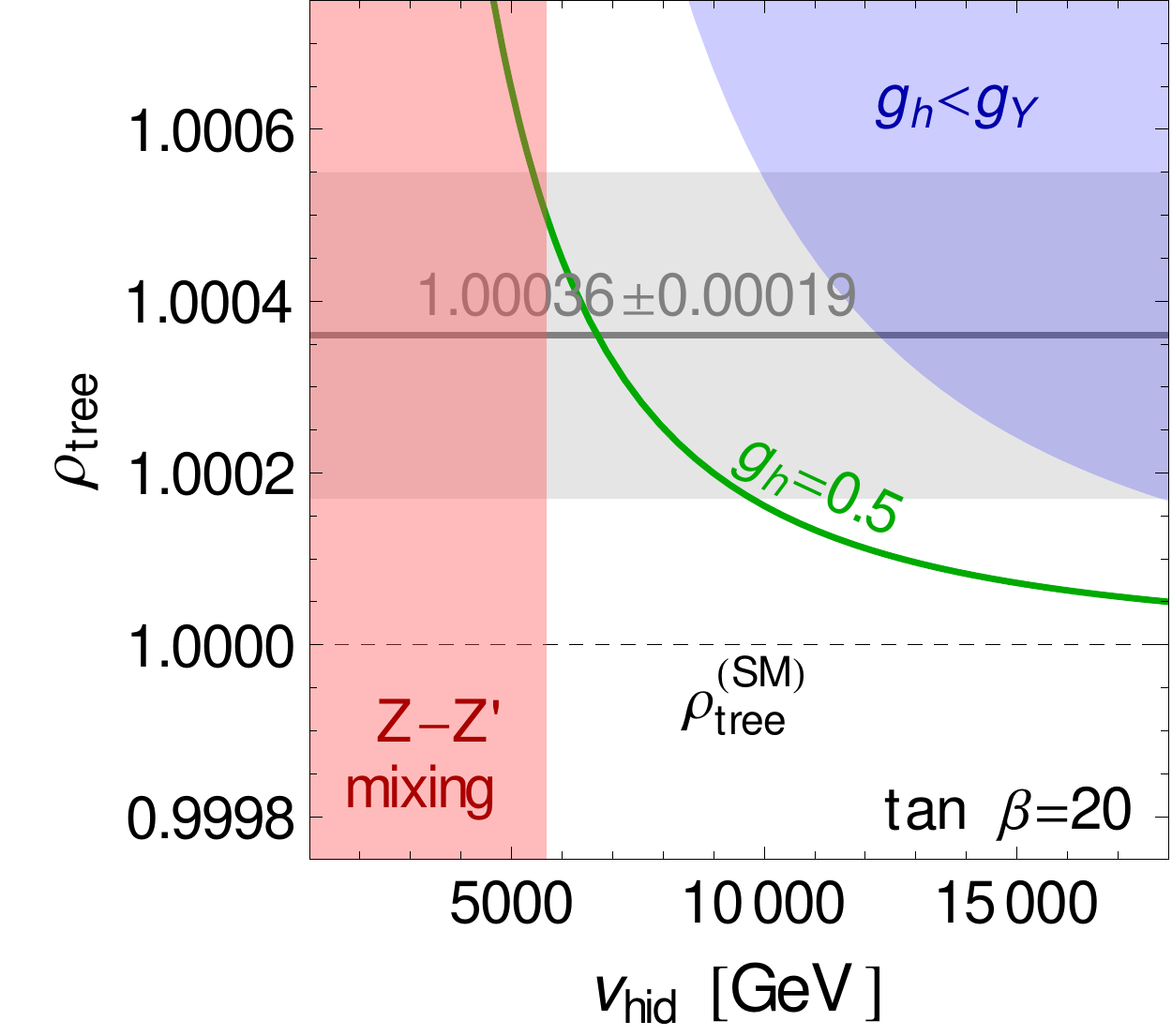}
\\
\includegraphics[scale=0.34]{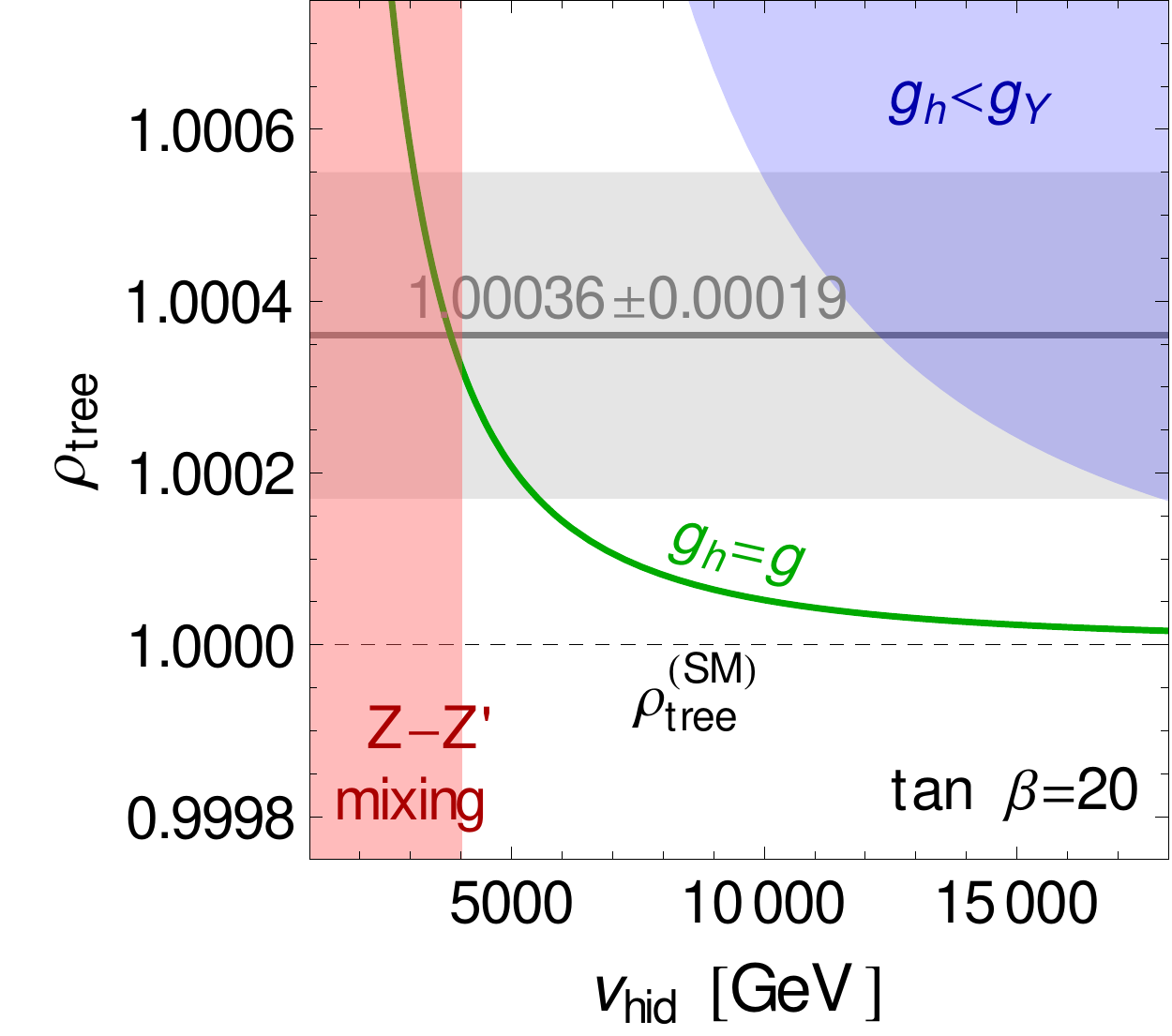}
\includegraphics[scale=0.34]{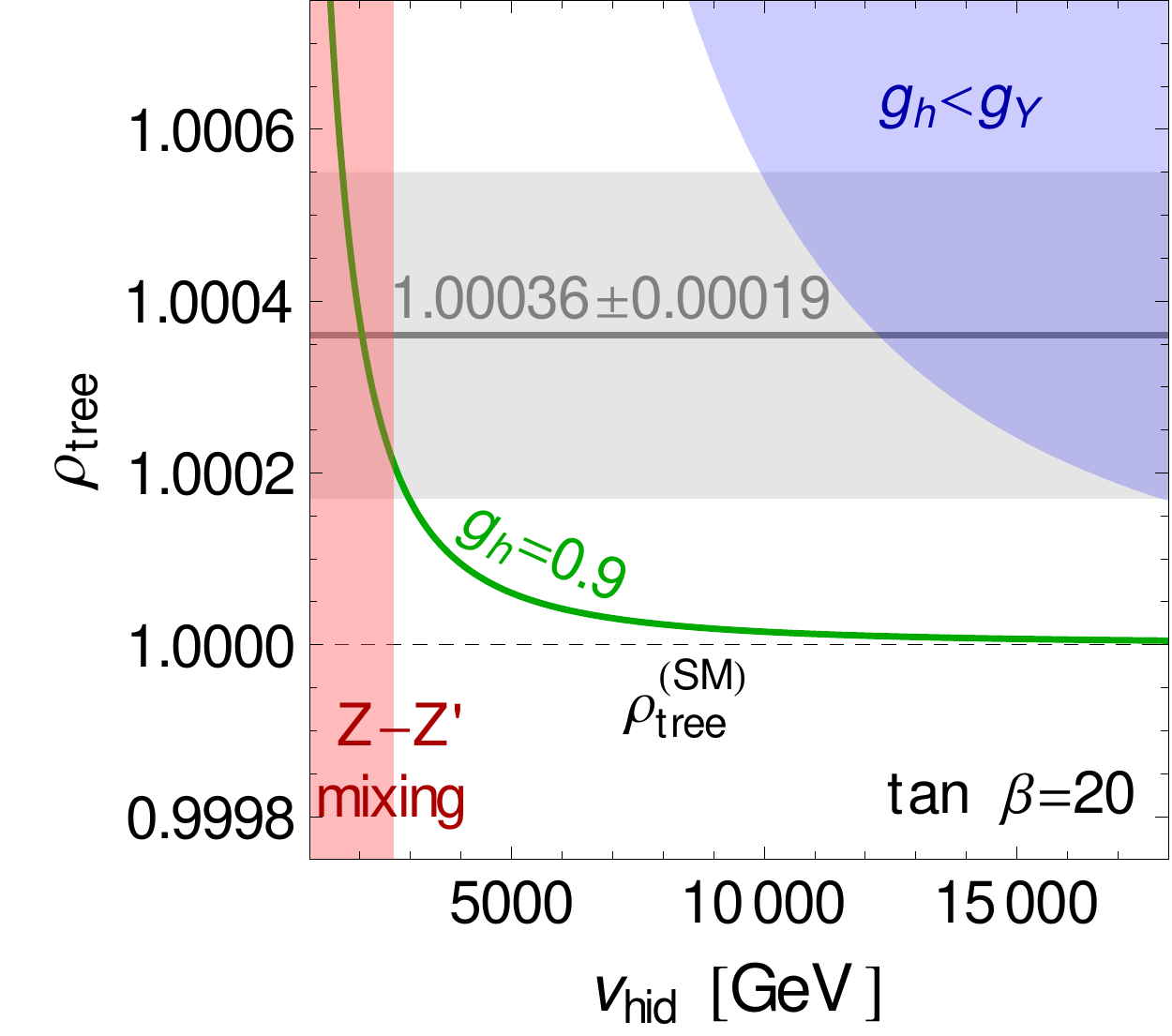}

\caption{Tree-level $\rho$ parameter variation with $v_{hid}$, at fixed $\tan \beta$, for several $g_{h}$ values (green solid lines in each panel). Latest fit and $1-\sigma$ band are shown in gray. Neither of $g_{h}$ of $g_{x}$ is allowed to take values below $g_{Y}$ (blue region). The red-shaded region, coming the bound from $Z-Z'$ mixing (in terms of $v_{hid}$) is excluded for each choice of $g_{h}$.}
\label{fig:rhoTree}
\end{figure}

The electroweak precision parameter $\rho$, on the other hand, gives some very interesting constraints at tree level due to the contribution from the vev of $H_{hid}$. Since the bidoublet $\mathcal{B}$ respects the custodial symmetry of the potential, only $v_{hid}$ contributes to $\rho$ in a nontrivial way. Denoting with $g$, $g_h$, $g_x$, and $g_Y$ the respective gauge coupling strengths of $SU(2)_w$, $SU(2)_{hid}$, $U(1)_X$, and $U(1)_Y$, we can write at tree-level 
\begin{equation}
\rho_{\text{tree}}=\dfrac{M_{W}^{2}}{c_{W}^{2}M_{Z}^{2}} =\dfrac{M_{W,\text{cust}}^{2}-\delta M_{W}^{2}}{c_{W}^{2}(M_{Z,\text{cust}}^{2}-\delta M_{Z}^{2})}~,
\end{equation}
where 
\begin{equation}
M^2_{W,\text{cust}} = g^2 v^2/4, \ \ M^2_{Z,\text{cust}} = (g^2 + g_Y^2) v^2/4,
\end{equation}
and
\begin{eqnarray}
c_{W} &=& \dfrac{g\sqrt{g_{h}^{2}+g_{x}^{2}}}{\sqrt{g^{2}(g_{h}^{2}+g_{x}^{2})+g_{h}^{2}g_{x}^{2}}} ,\\
g_{x}(g_{h}) &=& \dfrac{g_{Y}g_{h}}{\sqrt{g_{h}^{2}-g_{Y}^{2}}}~.\label{couplings-relation}
\end{eqnarray}
At order $O(v^{4}/v_{hid}^{2})$ we obtain
\begin{align}
\delta M_{W}^{2} &= \left( \dfrac{g^{2}\sin^{2}{(2\beta_{b})}}{4} \right)\dfrac{v_{b}^{4}}{v_{hid}^{2}}~, \label{eq:deltaW} \\
\delta M_{Z}^{2} &= \dfrac{ \bigl[ g_{x}^{2}g_{h}^{2}+g^{2}(g_{h}^{2}+g_{x}^{2})\bigr] }{4(g_{h}^{2}+g_{x}^{2})^{3}}\dfrac{ (g_{x}^{2}v_{w}^{2}-g_{h}^{2}v_{b}^{2})^{2} }{ v_{hid}^{2} }~.\label{eq:deltaZ}
\end{align}
Notice that both $\delta M_{W,Z}^{2}$ have fixed sign irrespective of $\tan \beta_{b}$ or the relative sizes between $g_{h}$ and $g_{x}$, thus the leading terms in an expansion in $v^{4}/v_{hid}^{2}$ carry opposite sign,
\begin{equation}
\rho_{\text{tree}}=1-\dfrac{\delta M_{W}^{2}}{M_{W,\text{cust}}^{2}}+\dfrac{\delta M_{Z}^{2}}{M_{Z,\text{cust}}^{2}}+O(v^{4}/v_{hid}^{4})~.\label{eq:rhoTaylor}
\end{equation}
Note that by being proportional to $v_{b}^{4}/v_{hid}^{2}$, $\delta M_{W}^{2}$ is suppressed at large $\tan \beta$ compared to $\delta M_{Z}^{2}$, which goes as $v_{w}^{4}/v_{hid}^{2}$. This results in a $\rho_{\text{tree}}>1$ that monotonically decreases to 1 as $v_{hid}$ increases (as expected). In order to show whether these tree-level contributions make $\rho_{\text{tree}}$ fall within the current best fit, $\rho^{(\text{exp})}=1.00036\pm0.00019$ \cite{Erler:2017vaq}, we show in Fig. \ref{fig:rhoTree} the $\rho_{\text{tree}}$ of Eq. (\ref{eq:rhoTaylor}) as a function of $v_{hid}$ for various $g_{h}$ at fixed $\tan \beta=20$.

Looking at Eq.\eqref{couplings-relation} we see that $g_{h}$ and $g_{x}$ must satisfy $g_{h},g_{x}>g_{Y}$, which rules out the blue region shown in each panel. This, combined with the fact that both bounds from the $\rho$ best fit lie above $1$, curiously implies an upper bound for $v_{hid}$ (when $g_h=0.37$) that translates in to a maximum value for the $W'^{\pm}$ mass, $M_{W}'\lesssim 3.2\tev$, as can be seen in the upper-left panel in Fig.~\ref{fig:rhoTree} (of course, associated to the $1 - \sigma$ band). As the chosen value for $g_h$ decreases, the allowed ranges for $v_{hid}$ (set by the crossings of the green contours with the gray band endpoints) decrease both in size and in value until they hit a bound obtained from $Z-Z'$ mixing (red shade in Fig. \ref{fig:rhoTree}), which reads $(\mathcal{M}_{Z}^{2})_{2,3}/(M_{Z'}^{2}-M_{Z}^{2})\lesssim 10^{-3}$ \cite{Patrignani:2016xqp} and that translates into $M'_W \sim 1\tev$ (lower-right panel). 

It is worthwile noting that even though these values for $M'_W$ fall below current constraints from general $W'^{\pm}$ searches ($M_{W'}>3-5\tev$ in \cite{Aaboud:2017efa},\cite{Khachatryan:2016jww}), our $W'^{\pm}$ evades standard production from and decay to SM states by carrying a reduced coupling to them. This is so because in $pp \to W'^{\pm}$ the SM quarks only couple to $W'^{\pm}$ through $W$-mixing of size $O(v_{b}^{2}/M_{W'}^2)$. Since the largest $W'$ couplings in the scalar sector are to mostly-bidoublet states (with $\ell'^{\pm}\nu$ assumed kinematically inaccesible) the partial fractions arrange hierarchically as
\begin{align}
\text{BR}(W'^{\pm}\to \mathcal{B}\mathcal{B})
&\gg~\text{BR}(W'^{\pm}\to \mathcal{B}~\text{SM}) \notag \\
&\gg~\text{BR}(W'^{\pm}\to \text{SM}~\text{SM}) \label{eq:Wdecays}
\end{align}
where $\mathcal{B}=H_{2,3}^{0},H_{1,2}^{\pm},A_{1,2}^{0}$  and $\text{SM}=h,W,Z$. Standard $W'$ searches based on $Wh$ and $WZ$ thus face large suppressing factors and are expected to be safe from current $W$-mixing constraints \cite{PhysRevD.98.030001}. On the other hand, the $W'$ can decay promptly ($\Gamma_{W'}\sim$ tens of $\text{GeV}$) thanks to the set of (large) BRs of  $\mathcal{B}\mathcal{B}$ pairs. However, although these $\mathcal{B}$'s decay first to pairs of $h,W,Z$ with not-so-small fractions, the unstability of the SM bosons implies that these unsuppressed $W'$ channels will contain large-multiplicity final states at detector-level that may require dedicated analyses.

In order to numerically exemplify the claims above, Table \ref{table:WBR} lists the largest BRs for $W'$ and its main decay products at $v_{h}=10.0\text{ TeV}$, $g_{h}=g_{2}\approx 0.66$, $\tan{\beta}=20.0$ supplemented with the scalar potential benchmark at the end of \ref{app:VHiggs}. At such point, $M_{W'}=3.3\text{ TeV}$ and $\Gamma_{W'}=9.5\text{ GeV}$, with bidoublet-like state masses $180-230\text{ GeV}$. Furthermore, by itself the factor $\text{BR}(W'^{\pm}\to q_{i}\bar{q}_{j})$ that enters into $\sigma(pp\to W'^{\pm})$ goes as $\sim 10^{-15}$, indeed diminishing the $W'^{\pm}$ production.

\begin{table}
	\begin{tabular}{|c|c||c|c|}
	\hline
	{\rm{$W'^{\pm}$ decays}} & BR & $\mathcal{B}$ decays & BR \\
	\hline \hline
	$H_{1}^{\pm}A_{1}^{0}$ & $\sim 25\%$ & $H_{1,2}^{\pm}\to t\bar{b}$ & $\sim 99\%$  \\
	$H_{2}^{\pm}A_{2}^{0}$ & $\sim 24\%$ & $A_{1}^{0}\to gg$ & $\sim 49\%$ \\
	$H_{1}^{\pm}H_{2}^{0}$ & $\sim 25\%$ & $A_{1}^{0}\to b\bar{b}$ & $\sim 42\%$ \\
	$H_{2}^{\pm}H_{3}^{0}$ & $\sim 25\%$ & $A_{2}^{0}\to hZ$ & $\sim 68\%$  \\
	$A_{1}^{0}W^{\pm}$ & $\sim 10^{-4}$ & $A_{2}^{0}\to b\bar{b}$ & $\sim 27\%$ \\
	$H_{1}^{\pm}Z$        & $\sim 10^{-4}$ & $H_{2,3}^{0}\to WW$ & $\sim 71\%$ \\
	$W^{\pm}Z$              & $\sim 10^{-10}$ & $H_{2,3}^{0}\to ZZ$ & $\sim 16\%$  \\
	$W^{\pm}h$              & $\sim 10^{-10}$ & & \\
	\hline
	\end{tabular}
	\caption{Main $W'^{\pm}$ branching fractions and a few representative subdominant ones. Also shown are the BRs of the mostly-bidoublet daughters of the main $W'^{\pm}$ modes.}
	\label{table:WBR}
\end{table}

Thus, this model gives the possibility of having a {\it relatively} light set of gauge bosons so far undetected that connect the two fermion sectors providing a way for electrically would-be stable particles to decay to SM fermion states. This feature is what motivates the adjective {\it hidden} for the extended gauge sector of our model.

A one-loop analysis will be soon presented for this case as well, where we also expect contributions to be under control due to the reduced nature of the couplings to SM fields. Another interesting loop-level process subject to possible constraints is the one associated to the diphoton strength signal, defined by the ratio $\mu_{\gamma \gamma}\equiv \sigma/\sigma_{\text{SM}}$ of cross sections of production of $h_{\text{SM}}$ that decays into $\gamma \gamma$. Deviations from the SM prediction $\mu_{\gamma \gamma}=1$ may serve as indicators of non-SM degrees of freedom, and since the latest $36\text{ fb}^{-1}$, $\sqrt{s}=13\tev$ ATLAS \cite{Aaboud:2018xdt} and CMS \cite{Sirunyan:2018ouh} results quote, respectively, $\mu_{\gamma \gamma}=0.99\pm0.14$ and $\mu_{\gamma \gamma}=1.18_{-0.14}^{+0.17}$, the current measurements leave room for contributions from the extra states in the model ($W'$, $q'$ and the charged scalars). To see that, consider the two charged scalars $H_{1,2}^{\pm}$ described in Sec. \ref{sec:model}.  For $v_{hid}$ fixed, the $H_{1,2}^{\pm}$ masses are governed by the sizes of vevs $v_{b1},v_{b2}$ and the parameters $\mu_{HBH},~\widetilde{\mu}'_{HBH}$ that mix the bidoublet with the doublets. Then, the $H_{1,2}^{\pm}$ can modify $\mu_{\gamma \gamma}$ from the SM prediction provided their masses are light enough  and their couplings to the SM-like Higgs are non-negligible. The $W'$, on the other hand, will be sufficiently decoupled to contribute for a $v_{hid}$ of several TeV. This also holds true for the hidden quarks $q'$ (whose mass is only sensitive to $v_{hid}$) provided their Yukawas $\mathcal{Z}_{ij}$ are taken of order 1. 

\begin{figure}[h!]
\centering
\includegraphics[scale=0.4]{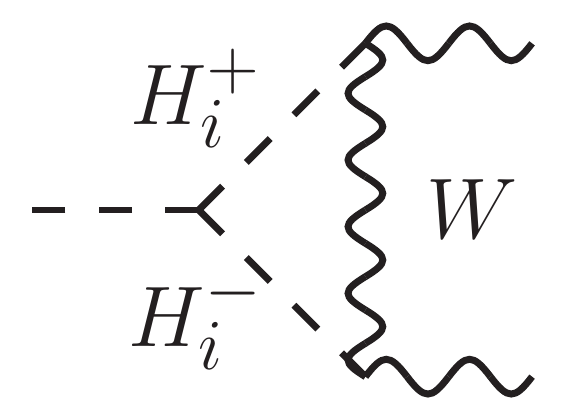}
\hspace{2.5mm}
\includegraphics[scale=0.4]{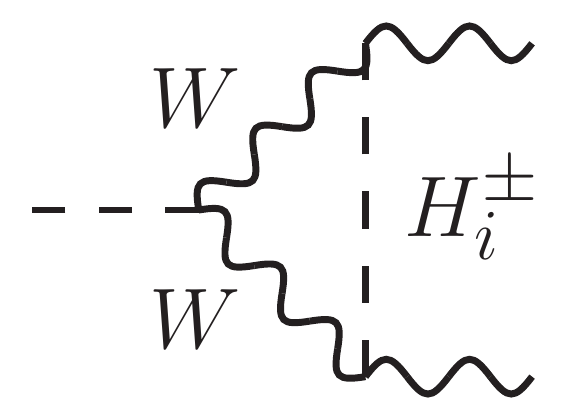}
\hspace{2.5mm}
\includegraphics[scale=0.4]{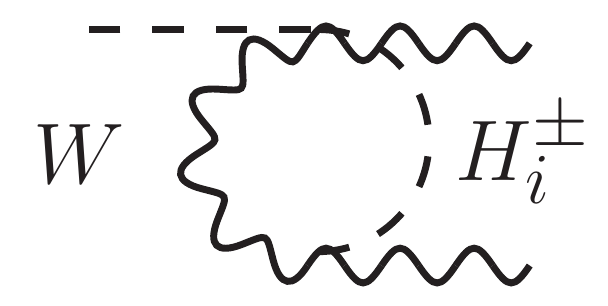}
\includegraphics[scale=0.4]{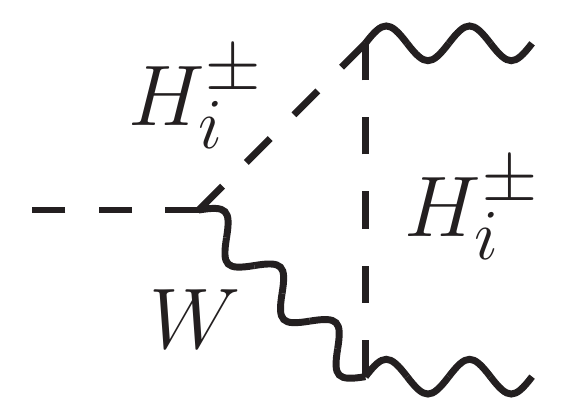}
\hspace{2.5mm}
\includegraphics[scale=0.4]{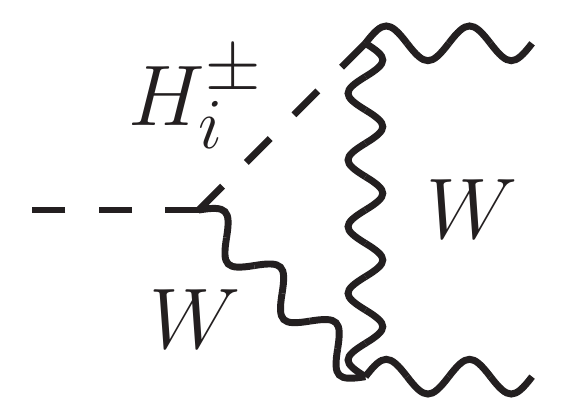}
\caption{Nonvanishing $h_{\text{SM}}\rightarrow \gamma \gamma$ modes involving multiparticle vertices with the bidoublet-like charged scalars $H_{i}^{\pm}$.}
\label{fig:hyyMulti}
\end{figure}

Incorporating non-negligible $H_{1,2}^{\pm}$ contributions of course involves determining the extent to which other couplings present in the SM amplitudes of $h_{\text{SM}}\rightarrow \gamma \gamma$ (those of the $W$ and $t$) are modified in our model, however, we would like to point out an interesting feature within our set up: while standard expressions for the total $\Gamma(h\rightarrow \gamma \gamma)$ incorporating new charged scalars exist (see \cite{Carena:2012xa}), these include only the usual triangle and bubble diagrams made solely of $H_{i}^{\pm}$ lines. For general mixing angles there are physical-basis, nonzero multiparticle vertices $\gamma W^{\pm}H^{\mp}$, $h_{\text{SM}}W^{\pm}H^{\mp}$ and $h_{\text{SM}}\gamma W^{\pm}H^{\mp}$ that generate the additional diagrams in Fig. \ref{fig:hyyMulti}. This situation is analogous of the one encountered in 2HDMs outside the $(\beta-\alpha)=\pi/2$ (\textit{alignment}) regime, where the coupling $h_{\text{SM}}W^{\pm}H^{\mp}$ no longer vanishes\footnote{Easier to see in the \textit{Higgs basis}, where the doublet orthogonal to the one with the Goldstones and the whole $v$ contains $H^{+}$ in the upper component and $-h_{\text{SM}}\cos{(\beta-\alpha)}$ in the lower component. Then a $W$ connects $H^{+}$ and $h_{\text{SM}}$.}.

Before closing the present discussion on the diphoton, let us mention that a similar scenario with two doublets and a $X=0$ bidoublet has been analyzed in Ref. \cite{Ashry:2013loa} in the context of a Left-Right model. Their set up is devoid of the amplitudes in Fig. \ref{fig:hyyMulti} thanks to the presence of an extra discrete symmetry which also forbids $\widetilde{W}-\widetilde{W}'$ mixing. Back to our model, a proper analysis of the modification to $h\rightarrow \gamma \gamma$ must not only incorporate the full set of diagrams and the rather numerous set of parameters in the scalar potential, but also do it in a way that ensures its boundedness from below.

These are the salient features that the present model offers. It is important to mention that we are also interested in future explorations about providing dark matter candidates and full fledged flavor models within this setting. There are also variants of the model presented in this paper that we are currently investigating, such as the case where the extended gauge group is $SU(3)_C \times SU(2)_w \times U(1)_Y \times SU(2)_{hid}$,  namely a case where $X=Y$(and $X(H_{hid})=0$). This leads to the peculiar result that there are two neutral (color singlet) fermions.

Another direction worth exploring in order to constraint $v_{hid}$ is the role of the $\nu_{R}$s in the context of BBN and CMB bounds. Being Dirac, these contribute to the total radiation density and could alter $\Delta N_{\text{eff}}$ in a non-negligible way via amplitudes that populate the $\nu_{R}$ number density \cite{Zhang:2015wua,Chacko:2016hvu,Borah:2017leo}, for example the decays $\ell'^{\pm}\to W'^{\pm}\nu_{R}$, $Z'\to \nu_{R}\nu_{R}$ and the scattering processes $W'^{+}W'^{-}\to \nu_{R}\nu_{R}$, $\ell'^{+}\ell'^{-}\to \nu_{R}\nu_{R}$. We hope to explore (and motivate others) these ideas in our ongoing investigations.

\section{Conclusions}
\label{sec:conclusion}
We have presented a set up where new generations of fermions and gauge interactions are incorporated into a model that is consistent with current experimental results. The model has gauge symmetry $SU(2)_C \times SU(2)_w \times SU(2)_{hid} \times U(1)_X$, three additional generations of fermions with {\it flipped} charges and chiralities, and three scalar multiplets (two doublets and a bidoublet). The set up uses the following convention for additional fermions and scalars: for any field that we add that has a counterpart within the SM, its $X$-charge (and chirality in the case of fermions) will be {\it flipped}. If additional fields that do not have a counterpart in the SM are needed, they will be $U(1)_X$-neutral.  These considerations lead to interesting phenomenological consequences such as the presence of relatively light {\it hidden} gauge bosons within the mass range of a few TeV, potential candidates for dark matter, and the natural presence of a Lepton number conserving dimension-five operator for Dirac neutrino masses. A peculiar result consists on the model imposing an upper limit to the mass of the additional gauge bosons of $M'_W \leq 3.2 \tev$. This upper bound comes from a combination of the conditions on the gauge couplings given by the symmetry breaking pattern and the contribution to the $\rho$ parameter at tree level. In order to motivate further exploration of this set up, we have outlined a couple of directions into the model's rich phenomenology.

\section*{Acknowledgements}
\label{sec:ack}
The work of A.A. was supported by CONACYT project CB-2015-01/257655 (M\'exico). C. A. thanks C\'{e}sar Bonilla for helpful discussions regarding the SARAH implementation.

\appendix
%******************** app:anomI ********************
\section{Anomaly coefficients}
\label{app:anomI}

Start with the charge-conjugates of the RH-chiral hidden quarks and leptons
\begin{equation}
Q_{R}^{'c}\sim(\boldsymbol{\bar{3}},\boldsymbol{1},\boldsymbol{2},+1/6),~~~~~R^{'c}\sim (\boldsymbol{1},\boldsymbol{1},\boldsymbol{2},-1/2)
\end{equation}
where generation indices are temporarily suppressed. There are new contributions $A^{\text{new}}(3,3,1)$, $A^{\text{new}}(J,J,1)$ and $A^{\text{new}}(1,1,1)$ to the SM coefficients, plus an entirely new coefficient $A(2_{hid},2_{hid},1)$ with no SM counterpart. For the former, with $N_{\text{gen}}$ the number of generations,
\begin{align}
A^{\text{new}}(3,3,1)
=N_{\text{gen}}&\bigl[ 2(+1/6)+(-2/3)+(+1/3) \bigr] \notag \\
&=0, \\
A^{\text{new}}(J,J,1)
=N_{\text{gen}}&\bigl[ 3\{ 2(+1/6)+(-2/3)+(+1/3) \} \notag \\
&+ \{ 2(-1/2)+(+1) \} \bigr]=0, \\
& \notag \\
A^{\text{new}}(1,1,1)
=N_{\text{gen}}&\bigl[ 3\{ 2(+1/6)^{3}+(-2/3)^{3}+(+1/3)^{3} \} \notag \\
&+ \{ 2(-1/2)^{3}+(+1)^{3} \} \bigr]=0.
\end{align}
The new coefficient arises from triangle diagrams with two $SU(2)_{hid}$ gauge bosons with $X^{\mu}$,
\begin{equation}
A(2_{hid},2_{hid},1)=N_{\text{gen}}\bigl[ 3\{ +1/6\}+(-1/2) \bigr]=0.
\end{equation}

\section{Bidoublet potential}
\label{app:VHiggs}
The potential $V_{\mathcal{B}}(H,H_{hid})$ written in Eq.(\ref{eq:Vshort}) is explicitly expanded below
\begin{eqnarray}\label{eq:fullVBH}
&  &V_{\mathcal{B}}(H,H_{hid}) \notag\\
& = &
 -\mu_{B}^{2}\text{Tr}\bigl[ \mathcal{B}^{\dag}\mathcal{B} \bigr]-\widetilde{\mu}_{B}^{2}\bigl( \text{Tr}\bigl[ \tilde{\mathcal{B}}\mathcal{B}^{\dag}]+\text{Tr}\bigl[ \tilde{\mathcal{B}}^{\dag}\mathcal{B} \bigr] \bigr)  \notag \\
&+& \lambda_{B}^{(1)}\text{Tr}\bigl[ \mathcal{B}^{\dag}\mathcal{B} \bigr]^{2}+\lambda_{B}^{(2)}\biggl( \text{Tr}\bigl[ \widetilde{\mathcal{B}}\mathcal{B}^{\dag} \bigr]^{2}+\text{Tr}\bigl[ \widetilde{\mathcal{B}}^{\dag}\mathcal{B} \bigr]^{2}  \biggr) \notag \\
&+&\lambda_{B}^{(3)}\text{Tr}\bigl[ \widetilde{\mathcal{B}}\mathcal{B}^{\dag} \bigr]\text{Tr}\bigl[ \widetilde{\mathcal{B}}^{\dag}\mathcal{B} \bigr] \notag \\
&+& \lambda_{B}^{(4)}\text{Tr}\bigl[ \mathcal{B}^{\dag}\mathcal{B} \bigr]\biggl( \text{Tr}\bigl[ \widetilde{\mathcal{B}}^{\dag}\mathcal{B} \bigr]+\text{Tr}\bigr[ \widetilde{\mathcal{B}}\mathcal{B}^{\dag} \bigr] \biggr) \notag \\
&+& \lambda_{BH}\text{Tr}\bigl[ \mathcal{B}^{\dag}\mathcal{B} \bigr]H^{\dag}H+\lambda'_{BH}\bigl[ \mathcal{B}^{\dag}\mathcal{B} \bigr]H_{hid}^{\dag}H_{hid} \notag \\
&+& \widetilde{\lambda}_{BH}\biggl( \text{Tr}\bigl[ \tilde{\mathcal{B}}\mathcal{B}^{\dag}]+\text{Tr}\bigl[ \tilde{\mathcal{B}}^{\dag}\mathcal{B} \bigr] \biggr)H^{\dag}H \notag \\
&+& \widetilde{\lambda}'_{BH}\biggl( \text{Tr}\bigl[ \tilde{\mathcal{B}}\mathcal{B}^{\dag}]+\text{Tr}\bigl[ \tilde{\mathcal{B}}^{\dag}\mathcal{B} \bigr] \biggr)H_{hid}^{\dag}H_{hid} \notag \\
&+& \biggl( \mu_{HBH}H^{\dag}\mathcal{B}\widetilde{H}_{hid}+\widetilde{\mu}'_{HBH}H^{\dag}\widetilde{\mathcal{B}}\widetilde{H}_{hid}+\text{H.c.} \biggr)~. \notag
\end{eqnarray}
This is the most general gauge-invariant potential in two doublets and one $X=0$ bidoublet \cite{Borah:2010zq}. Its non-triviality is due to the various possible contractions not only of $\mathcal{B}$, but of $\widetilde{\mathcal{B}}\equiv \sigma_{2}\mathcal{B}^{*}\sigma_{2}$ as well. All $\lambda$'s are dimensionless, and $[\mu_{HBH}]=[\widetilde{\mu}'_{HBH}]=+1$. Four minimization conditions, corresponding for instance to $v_{w},v_{b1},v_{b2}$ and $v_{hid}$, can be extracted with the help of the model generator SARAH \cite{Staub:2008uz}. The benchmark set used in \ref{table:WBR} is
\begin{align*}
&\lambda=0.15,~~~\lambda'=0.1,~~~\lambda_{HH}=-0.1~,\\
&\lambda_{B}^{(1-4)}=(0.1,0.1,-0.5,-0.1)~, \\
&\lambda_{BH}=\lambda'_{BH}=0.1,~~~\widetilde{\lambda}_{BH}=\widetilde{\lambda}'_{BH}=0.01,~ \\
&\mu_{HBH}=0.25\text{ GeV},~~~\widetilde{\mu}_{BHB}=-0.05\text{ GeV}
\end{align*}
with the BRs running through the SPheno \cite{Porod:2003um} spectrum generator.

\bibliographystyle{elsarticle-num}
\bibliography{allReferences.bib}

\end{document}